\documentclass[aip,apm,twocolumn,showpacs,superscriptaddress, preprintnumbers,amsmath,amssymb,a4]{revtex4-1}
\usepackage{graphicx}
\draft 

\begin{document}


\title{Intrinsic carrier mobility of multi-layered MoS$_2$ field-effect transistors on SiO$_2$} 



\author{N. R. Pradhan}
\email[]{pradhan@magnet.fsu.edu}
\affiliation{National High Magnetic Field Laboratory, Florida
State University, Tallahassee-FL 32310, USA}
\author{D. Rhodes}
\affiliation{National High Magnetic Field Laboratory, Florida
State University, Tallahassee-FL 32310, USA}
\author{Q. Zhang}
\affiliation{National High Magnetic Field Laboratory, Florida
State University, Tallahassee-FL 32310, USA}
\author{S. Talapatra}
\affiliation{Physics Department, Sourthern Illinois University, Carbondale-IL 62901-4401, USA}
\author{M. Terrones}
\affiliation{Department of Physics, Department of Materials Science and Engineering and Materials Research Institute. The Pennsylvania State University, University Park, PA 16802, USA}
\author{P. M. Ajayan}
\affiliation{Department of Mechanical Engineering and Materials Science, Rice University, Houston, TX 77005 USA}
\author{L. Balicas}
\email[]{balicas@magnet.fsu.edu}
\affiliation{National High Magnetic Field Laboratory, Florida
State University, Tallahassee-FL 32310, USA}


\date{\today}

\begin{abstract}
By fabricating and characterizing multi-layered MoS$_2$-based field-effect transistors in a four terminal configuration, we demonstrate that the two terminal-configurations tend to underestimate the carrier mobility $\mu$ due to the Schottky barriers at the contacts. For a back-gated two-terminal configuration we observe mobilities as high as 91 cm$^2$V$^{-1}$s$^{-1}$ which is considerably smaller than 306.5 cm$^2$V$^{-1}$s$^{-1}$ as extracted from the same device when using a four-terminal configuration. This indicates that the intrinsic mobility of MoS$_2$ on SiO$_2$ is significantly larger than the values previously reported, and provides a quantitative method to evaluate the charge transport through the contacts.
\end{abstract}


\maketitle 


Transition metal dichalcogenides are layered materials characterized by strong in-plane covalent bonding and weak inter-planar van der Waals coupling. Similarly to graphene, this weak interactions allow the exfoliation of these materials into two-dimensional layers with just a few or even a single layer thickness \cite{kis_review}. These compounds have been studied for decades, but recent advances in nanoscale characterization and device fabrication have opened up the potential for applications of thin, two-dimensional layers of dichalcogenides in nanoelectronics \cite{circuits1, circuits2} and in optoelectronics \cite{phototransistor}. Bulk compounds such as MoS$_2$, MoSe$_2$, WS$_2$ and WSe$_2$ exhibit indirect bandgaps in the order of 1.3 \={e}V which are expected to become direct bandgaps in single layers, thus making them excellent candidates for the development of photodetectors and electroluminescent devices.

Inspired by the extremely large carrier mobilities observed in graphene \cite{dean}, i.e. in excess of 100000 cm$^2$/Vs, much of the recent experimental effort is devoted to finding ways of increasing the carrier mobility in field-effect transistors (FETs) based on single- or few-layered dichalcogenides. For example, the deposition of a high-$\kappa$ dielectric layer (i.e. HfO$_2$) followed by a metallic top-gate onto exfoliated single-layer  MoS$_2$ on SiO$_2$ substrates was found to increase the mobility by nearly two orders of magnitude \cite{kis} to $\sim 200$ cm$^2$/Vs. For FETs based on 5 to 12 nm thick exfoliated MoS$_2$ single-crystals on 300 nm SiO$_2$ substrates \cite{peide}, the mobility is found to remain nearly independent of the length $\ell_c$ of the conduction channel for $0.5 \leqq \ell_c \leqq 2$ $\mu$m implying that by decreasing the thickness of the SiO$_2$ layer, it would in principle be possible to maintain high mobilities in devices having $\ell_c \sim 10$ {\AA}. Finally,
a combination of Sc contacts with Al$_2$O$_3$ as the dielectric used in a top gate configuration \cite{das} was shown to lead to mobilities as high as 700 cm$^2$/Vs at room temperature. Remarkably, the authors of Ref. \onlinecite{das} demonstrated the existence of sizeable Schottky barriers for virtually all metals used for the electrical contacts, even when the current-voltage characteristics display Ohmic behavior which is attributed to thermally assisted tunneling.

Sizeable Schottky barriers at the level of the contacts limit the current output of field-effect transistors which leads to lower, extrinsic values for the mobility of the charge carriers if their response is measured in a two-contact configuration. Here, we show by using a four-contact configuration which eliminates the role of the contacts, that the intrinsic carrier mobility of FETs based on multi-layered ($\sim$ 20 layers thick) MoS$_2$ on SiO$_2$ is nearly one order of magnitude higher than previously reported for back gated devices, i.e. $\simeq 300$ cm$^2$/Vs when compared to a few tenths of cm$^2$/Vs, see for example, Ref. \onlinecite{peide}. Thus, a simple comparison between the mobilities obtained for either configuration of electrical contacts, i.e. 4 versus 2 contacts provides a simple way to extract the intrinsic mobility of any given device and a way to evaluate the quality of the contacts at the metal-semiconductor interface, i.e. the higher and the closer are the values of the mobility measured in each configuration, the lower the height of the Schottky barrier.
\begin{figure*}[htb]
\begin{center}
\includegraphics[width = 11 cm]{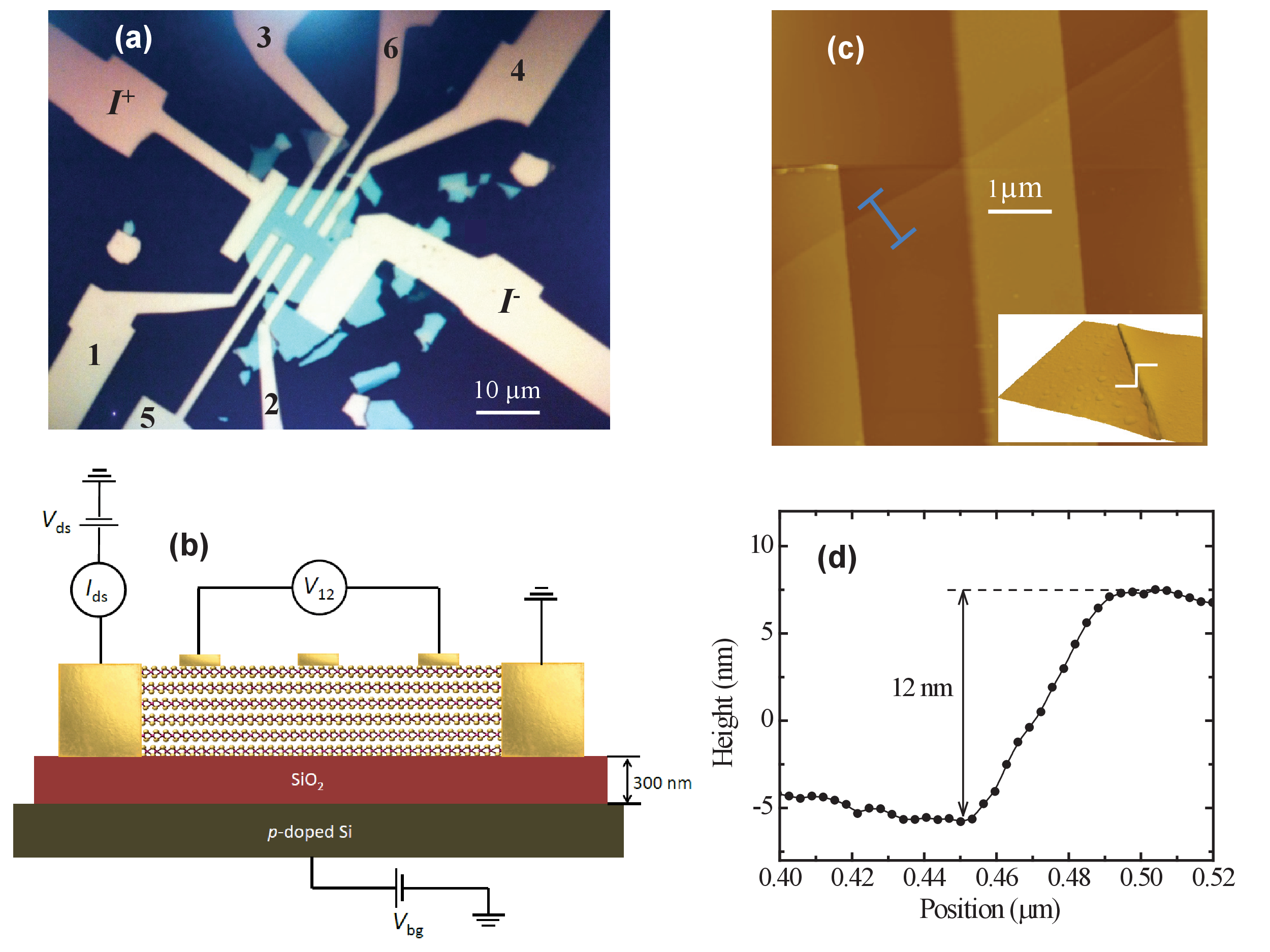}
\caption{ (a) Photograph of one of our multi-layered MoS$_2$ field-effect transistors as observed through an optical microscope. Gold pads correspond to voltage (pads labeled as 1, 2, 3, 4, 5 and 6) and current leads (pads labeled as $I^{+}$ and $I^{-}$). The distance $\ell_c$ between current leads is 11 $\mu$m, the separation $\ell_v$ between voltage leads 1 and 2, or 3 and 4 is 6.5 $\mu$m while the average width of the flake is 11.6 $\mu$m.
(b) Sketch of one of our devices when measured in a four terminal configuration. (c) Atomic force microscopy image (top view) across the edge of the device previously shown in Figs. (a) and (b), respectively. Inset: lateral AFM perspective of the edge of the device. Blue line and white lines indicates the line along which the height profile shown in (d) was measured. (d) Height profile across the edge of the device indicating a thickness of $\simeq$ 120 {\AA} or approximately 19 atomic layers. }
\end{center}
\end{figure*}

Single layers of MoS$_2$ were exfoliated from commercially available crystals of molybdenite (SPI Supplies Brand Moly Disulfide) using the scotch-tape micromechanical cleavage technique, and transferred onto \emph{p}-doped Si wafers covered with a 300 nm thick layer of SiO$_2$. Contacts were patterned by using standard e-beam lithography techniques. For making the electrical contacts 90 nm of Au was deposited onto a 4 nm layer of Ti via e-beam evaporation. After gold deposition the devices were annealed at $200$ $^{\circ}$C for $\sim$ 2 h in forming gas.  Atomic force microscopy (AFM) imaging was performed using the Asylum Research MFP-3D AFM. Electrical characterization was performed with a Keithley 2612A dual sourcemeter coupled to a Physical Parameter Measurement System.

Figure 1 (a) shows an optical microscopy image of one of our devices based on multi-layer MoS$_2$ in a 8 contacts configuration, i.e. 6 contacts which can be used for 4 point resistivity and Hall-effect measurements (which will not be discussed here) and 2 contacts for current injection. For the four terminal measurements shown here, the voltage was measured between gold leads 1 and 2 and also between 3 and 4. We checked that the results being displayed and discussed below are consistent with results obtained in another two devices, one revealing similar mobilities and a second one displaying lower mobilities, i.e. in the order of tenths of cm$^2$/Vs. Fig. 1 (b) displays a sketch of the field-effect transistor and of the configuration of measurements used. Fig. 1 (c) shows an AFM image around the edge of the FET. The thickness of the MoS$_2$ crystal was measured by scanning the AFM tip along the blue line (or along the white line in the inset). Fig. 1 (d) shows the resulting height profile when scanning the AFM tip from the SiO$_2$ substrate towards the surface of the MoS$_2$ crystal, it shows a step of approximately 12 nm, indicating that our device is composed of nearly 19 mono-layers.
\begin{figure}[htb]
\begin{center}
\includegraphics[width = 7 cm]{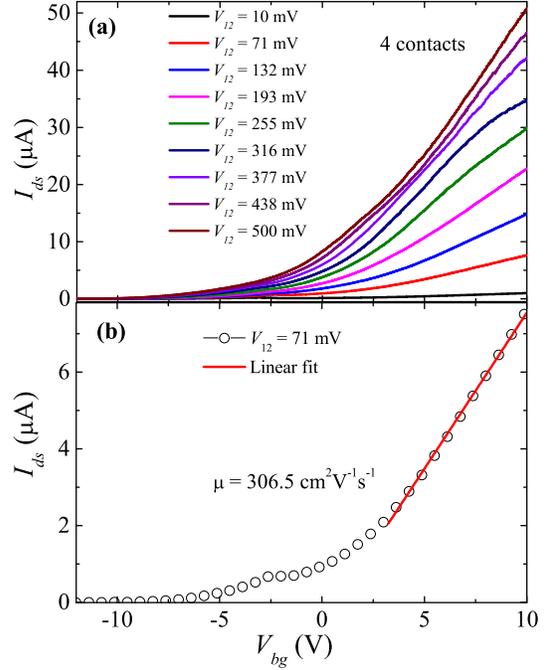}
\caption{(a) Electrical current $I_{ds}$ flowing through the current leads as a function of the back-gate voltage $V_{bg}$ for several constant values of the voltage $V_{12}$ across the voltage leads. (b) Based on Eq. (1), from the slope of $I_{ds}$ as a function of $V_{bg}$ acquired under a constant value $V_{12} = 71$ mV one extracts a carrier mobility $\mu = 306.5$ cm$^2$V$^{-1}$s$^{-1}$.}
\end{center}
\end{figure}

Figure 2 (a) shows the drain-current $I_{ds}$ as a function of back-gate voltage $V_{bg}$ for several constant values of the voltage $V_{12}$ between voltage leads. In agreement with all previous reports, MoS$_2$ shows unipolar response. Fig. 2 (b) shows $I_{ds}$ as a function of $V_{bg}$ for $V_{12} = 71$ mV, from the slope (red line corresponds to a linear fit) we extract the carrier mobility through the expression \cite{podzorov}:
\begin{equation}
\mu =\left[\frac{\ell_v}{wC_i} \right] \times \left[\left(d \left(\frac{I_{ds}-I_0}{V_{12}} \right)/dV_{bg} \right) \right]
\end{equation}
Where $C_i = 11.505 \times 10^{-9}$ F/cm$^2$ is the capacitance between the gate and the channel for a 300 nm layer of SiO$_2$ ($C_i =\varepsilon_0 \varepsilon_r/d$; $\varepsilon_r = 3.9$; $d =300$ nm), $\ell_v$ is the distance between the voltage leads 1 and 2, $w$ is the width of the channel, and $I_0$ is the current in the subthreshold regime \cite{podzorov2}.  Remarkably, the value of the extracted mobility is sizeable i.e. 306.5 cm$^2$/Vs and to the best of our knowledge, this is largest value for the mobility of a MoS$_2$ based FET on SiO$_2$ without the use of any ``dielectric engineering". This value is independent of the constant voltage $V_{12}$, i.e. for each trace in Fig. 2 (a) one systematically obtains $(305 \pm 30 )$ cm$^2$/Vs. This value is close to the calculated intrinsic mobility of MoS$_2$ based FETs on SiO$_2$, i.e. $\sim 410$ cm$^2$/Vs and which is limited by optical phonon scattering \cite{jacobsen}. It is larger than the value of 217 cm$^2$/Vs previously reported for single-layered MoS$_2$ using HfO$_2$ in a top gate configuration \cite{kis}.  We have performed the same measurements for different pairs of voltage leads in all devices finding variations of about 10 \% in the extracted values of the mobility in any given device.

\begin{figure}[htb]
\begin{center}
\includegraphics[width = 8 cm]{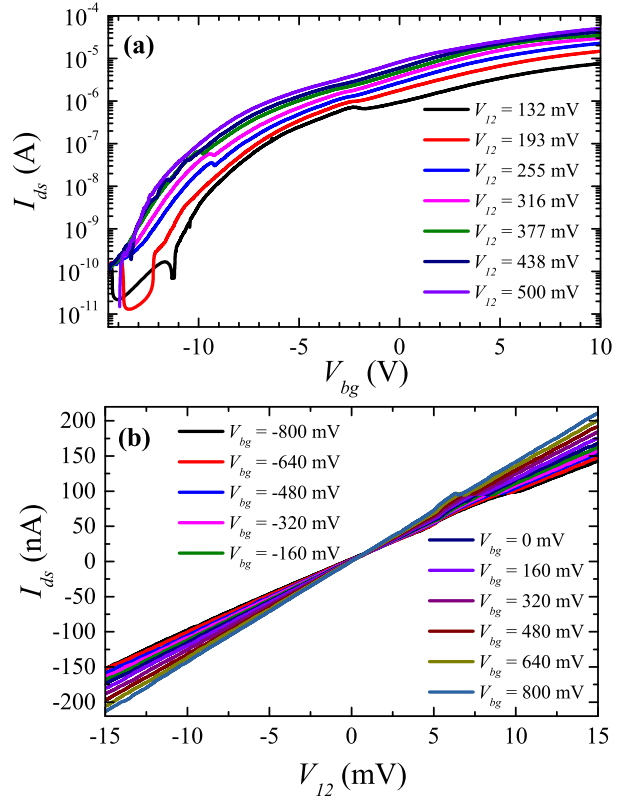}
\caption{(a) $I_{ds}$ in a logarithmic scale as a function of the back-gate voltage $V_{bg}$ for several constant values of $V_{12}$. Notice how $I_{ds}$ increases by $\simeq$ 6 orders of magnitude when $V_{bg}$ is swept from -14 to 10 V. (b) Current $I_{ds}$ as a function of the voltage $V_{12}$ for several values of constant back-gate voltage $V_{bg}$. Notice the nearly linear or Ohmic-like behavior. The four-terminal resistance of the flake $\simeq 87.7$ k$\Omega$ can be extracted from the inverse of the slope of the $V_{bg} = 0$ V curve.}
\end{center}
\end{figure}
\begin{figure}[htb]
\begin{center}
\includegraphics[width = 8 cm]{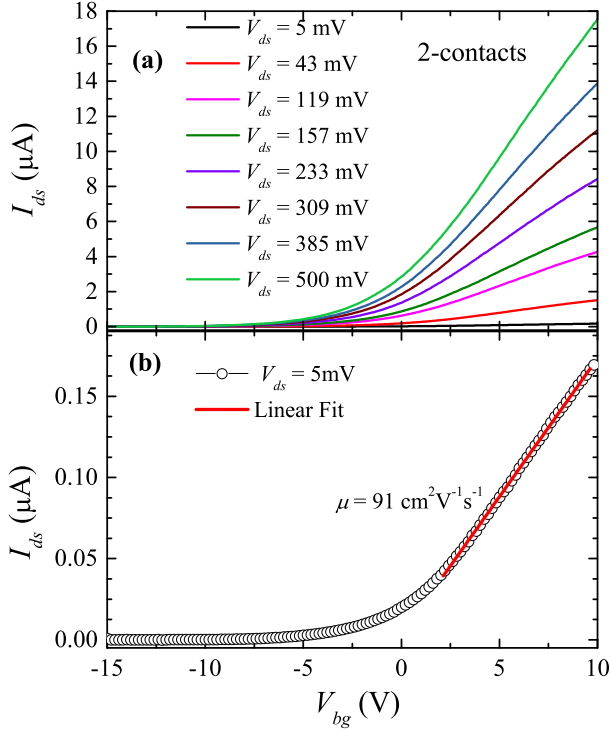}
\caption{(a) Electrical current $I_{ds}$ flowing through the current leads as a function of the back-gate voltage $V_{bg}$ for several constant values of the voltage $V_{ds}$ across the current leads. (b) Based on Eq. (2), from the slope of $I_{ds}$ as a function of $V_{bg}$ acquired under a constant value $V_{ds} = 5$ mV one extracts a carrier mobility $\mu = 91$ cm$^2$V$^{-1}$s$^{-1}$.}
\end{center}
\end{figure}

Figure 3 (a) shows the same $I_{ds}$ as a function of back-gate voltage $V_{bg}$ previously shown in Fig. 2 (a), but in a logarithmic scale. As seen, by scanning the back-gate voltage $V_{bg}$ from $\sim -15$ to 10 V, $I_{ds}$ increases by nearly six orders of magnitude. The subthreshold region is observed for $V_{bg} \lesssim -14$ V and \emph{not} for $V_{bg} \sim 0$ V thus indicating that our MoS$_2$ FETs are doped with electrons. Fig. 3 (b) shows the current $I_{ds}$ as a function of the voltage $V_{12}$ for different values of the back-gate voltage $V_{bg}$.
As expected, the $I_{ds}$ as function of $V_{12}$ displays linear or Ohmic behavior.

Figure 4 (a) shows the drain-current $I_{ds}$ as a function of back-gate voltage $V_{bg}$ for several constant values of the voltage $V_{ds}$ between current leads, or the FET $I-V$ characteristics for a two-contact configuration. Fig. 4 (b) shows $I_{ds}$ as a function of $V_{bg}$ for $V_{ds} = 5$ mV, from the slope (red line is a linear fit) we extract the carrier mobility through the usual expression:
\begin{equation}
\mu = \left(\frac{dI_{ds}}{dV_{bg}}\right) \times \left[\frac{\ell_c}{w V_{ds}C_i} \right]
\end{equation}
Where $C_i = 11.505 \times 10^{-9}$ F/cm$^2$ and $\ell_c$ is the distance between source and drain contacts.
Remarkably, even in a two contact configuration we obtain a carrier mobility of 91 cm$^2$/Vs, which is higher than the mobilities reported by other groups, e.g. Ref. \onlinecite{peide}, for multi-layered MoS$_2$ on SiO$_2$.
We have checked that this value is independent of the applied $V_{ds}$ voltage, one systematically obtains $(90 \pm 5 )$ cm$^2$/Vs. Surprisingly, this value is comparable to the one recently reported \cite{kim} for multilayered MoS$_2$ on a 50 nm thick Al$_2$O$_3$ layer, i.e. $\mu \sim 100$ cm$^2$/Vs. At first glance this observation would seem to challenge the notion of ``dielectric engineering" as a route for increasing the carrier mobility in dichalcogenide based FETs. As expected, the extracted mobility is considerably smaller, i.e. by factor $ \sim 3$, than the value extracted from the four-point measurement, which is a clear indication for the role played by the Schottky barriers at the level of the contacts.
\begin{figure}[htb]
\begin{center}
\includegraphics[width = 7 cm]{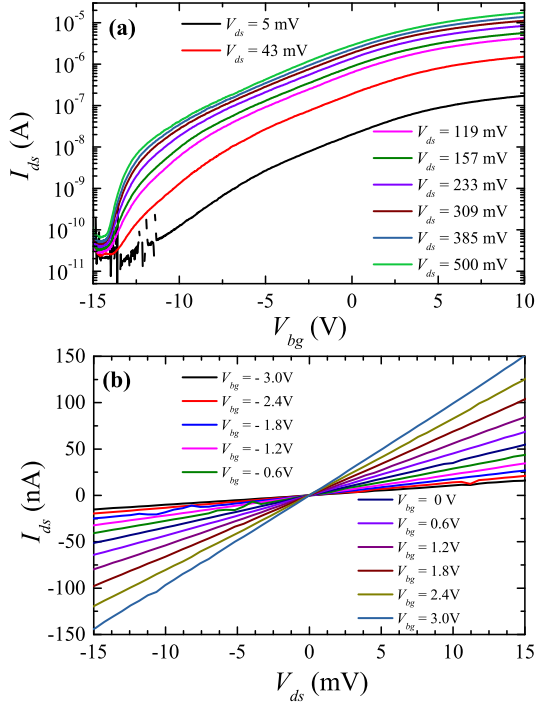}
\caption{(a) $I_{ds}$ in a logarithmic scale as a function of the back-gate voltage $V_{bg}$ for several constant values of $V_{ds}$. Notice how $I_{ds}$ increases by $\simeq$ 6 orders of magnitude when $V_{bg}$ is swept from -17.5 to 10 V. (b) Current $I_{ds}$ as a function of the voltage $V_{ds}$ for several values of constant back-gate voltage $V_{bg}$. Notice the Ohmic-like response. The two-terminal resistance of the flake $\sim 280$ k$\Omega$ can be extracted from the inverse of the slope of the $V_{bg} = 0$ V curve.}
\end{center}
\end{figure}

Figure 5 (a) shows the drain-current $I_{ds}$ in a logarithmic scale as a function of the back-gate voltage $V_{bg}$ for several constant values of the voltage $V_{ds}$. This is the same data set previously shown in Fig. 4 (a) in a linear scale. As seen, for a given value of $V_{bg}$ the amount of current $I_{ds}$ extracted from the device depends strongly on the applied voltage $V_{ds}$. $I_{ds}$ as function of $V_{bg}$ increases by only $\sim 4$ orders of magnitude when $V_{ds} \leq 10$ mV, but it is seen to increase by nearly 6 orders of magnitude when $V_{ds} = 500$ mV. This can be easily understood if one considers that the resistance of the contacts is strongly dependent on
the back gate bias as shown by Ref. \onlinecite{peide}; the resistance of the contacts decreases as the MoS$_2$ crystal is electrically doped under high gate bias, leading to smaller contact resistances. As discussed in Ref. \onlinecite{peide}, the gate voltage dependence of the resistance of the contacts can be attributed to i) the existence of a Schottky barrier at the metal/semiconductor interface since the gate voltage changes the tunneling efficiency due the bending of the band at the metal/semiconductor interface, and ii) to the electrical doping of the semiconductor. Remarkably, and despite the above observations, the $I_{ds}$ as a function of $V_{ds}$ characteristics displayed in Fig. 5 (b) shows a remarkably linear or Ohmic dependence, which at first glance would seem at odds with the existence of a sizeable Schottky barrier. However, and as argued in Ref. \onlinecite{das} the linear $I-V$ characteristics is misleading and would result from thermally assisted tunneling through the Schottky barriers. Finally, our room temperature 2- (Fig. 5 a) and 4-terminal (Fig. 3 a) measurements indicate that this transistor current on/off ratio exceeds $10^6$.

Therefore, we can conclude that carrier mobility in field-effect transistors based on few-layered transition-metal dichalcogenides, as reported by a number of groups, is strongly limited by non-Ohmic contacts. A problem that has yet to be circumvented. However, and as the present work indicates, a systematic comparison between 4- and 2-terminal configurations for a variety of metals, combinations of thereof, and treatments should lead to a solution to this important issue. Notice nevertheless, that a previous study on similar MoS$_2$ devices using a four-terminal configuration for the contacts revealed considerably smaller mobilities than the ones reported here. At the moment we do not have a clear physical explanation for this difference. Another serious mobility limiting factor as discussed in Ref. \onlinecite{ghatak}, is the role of the substrates which in the case of SiO$_2$ leads to charge localization. For example, for the device shown here we have seen a very modest increase in mobility, i.e. $\sim 10$ \%, from room temperature to 150 K indicating that optical phonons are not the only relevant scattering mechanism. Naively, one would have expected the suppression of phonon scattering and an effective increase in carrier mobility as the temperature is lowered and as is indeed seen for top gated devices \cite{kis2} using HfO$_2$. Therefore, it is quite likely that an atomically flat substrate, free from dangling bonds such as \emph{h}-BN, in combination with the proper electrical contacts, and the use of a high-$\kappa$ dielectric, might lead one day to unexpectedly high values for the mobility and concomitant on/off ratios in single- and few-layered transition metal dichalcogenides field-effect transistors.

This work is supported by the U.S. Army Research Office MURI grant W911NF-11-1-0362.
The NHMFL is supported by NSF through NSF-DMR-0084173 and the
State of Florida.



%
%

%


{}
\end{document}